# COMPARING ONLINE COMMUNITY STRUCTURE OF PATIENTS OF CHRONIC DISEASES

Hanuma Teja Maddali, Peter A. Gloor, Peter A. Margolis
MIT CCI, Cincinnati Children's Medical Center
Cambridge USA, Cincinnati USA
hmaddali@mit.edu, pgloor@mit.edu, Peter.Margolis@cchmc.org

## ABSTRACT

In this paper we compare the social network structure of people talking about Crohn's disease, Cystic Fibrosis, and Type 1 diabetes on Facebook and Twitter. We find that the Crohn's community's contributors are most emotional on Facebook and Twitter and most negative on Twitter, while the T1D community's communication network structure is most cohesive.

## INTRODUCTION

Social media has become a major means of communication for patients of chronic diseases; to stay in touch with each other, find support and learn about novel treatments to better cope with their illness. In particular, Facebook has become a major channel for community activation and peer group support. In earlier work we found that while patients of Crohn's disease actively participated on Facebook discussions by posting comments, they were surprisingly unconnected by not "friending" each other on Facebook (Gloor et al. 2010). In this project we build on this earlier work by extending the focus comparing Facebook pages on different chronic illnesses, and also looking at how patients and other stakeholders talk about the same chronic diseases on Twitter.

In this first analysis we focus on Facebook groups and Tweets about "Crohn's", "T1D" and "Cystic Fibrosis". The Condor 3.1 toolkit was used for network and content analysis.

## ANALYZING FACEBOOK PAGES

Facebook pages are an excellent source of ethnographic information since they are public and accessible through an API. Because they are usually the official pages of organizations, they are more likely to be moderated compared to Facebook groups, thus also containing less spam. We found 517 Facebook pages about cystic fibrosis, 275 groups about type 1 diabetes, and 587 groups about Crohn's disease.

Table 1 lists the key statistics of the top 4 Facebook pages for each of the three chronic diseases. Crohns is the most active, as it has gathered close to 10,000 messages in just seven months.

| Chronic Disease | Facebook Page Name | Number of Likes | Actors | Messages | Data time range |
|---|---|---|---|---|---|
| Cystic Fibrosis | Cystic Fibrosis Foundation | 188194 | 4302 | 5669 | Sep 23, 2011 to Nov 28, 2014 |
| | Cystic Fibrosis Trust | 64527 | | | |
| | CysticFibrosis.com | 17567 | | | |
| | Cystic Fibrosis Canada | 9080 | | | |
| Type 1 Diabetes | Type 1 Diabetes | 37643 | 3906 | 9217 | Jan 18, 2010 to Nov 28, 2014 |
| | Cure Type 1 Diabetes | 8636 | | | |
| | I hate Diabetes (type 1) | 6422 | | | |
| | Type 1 Diabetes Awareness | 4960 | | | |
| Crohns | Crohn's and Colitis UK | 90989 | 6143 | 9836 | May 1, 2014 to Nov 28, 2014 |
| | CCFA - Crohn's & Colitis Foundation of America | 89481 | | | |
| | Crohn's & Colitis Awareness | 30265 | | | |
| | Living with Crohn's Disease: Healthline | 17451 | | | |

Table 1: Statistics of top 4 Facebook pages for each of Crohn's, T1D and cystic fibrosis

It appears that the official patient organizations of Cystic Fibrosis and Crohn's are doing a better job activating patients on Facebook compared to T1D, as they are collecting between 90,000 and 190,000 likes, while the T1D top pages are "unofficial" pages with much lower numbers of likes.

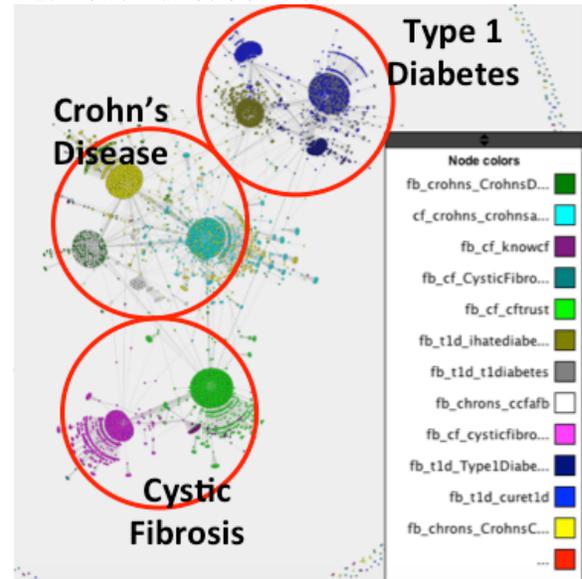

Figure 1: Combined network structure of three patient communities of top 4 FB pages

Figure 1 shows the network created by the "walls" of the Facebook pages, with a link between two actors if one responded to a wall post of another actor. The central node for each of the clusters in figure 1 is the

Facebook page. The connecting nodes between the pages are people posting on more than one Facebook page. As figure 1 shows, there are people posting about more than one disease, for instance both about Crohn's and Cystic Fibrosis. Figure 2 lists the key statistics.

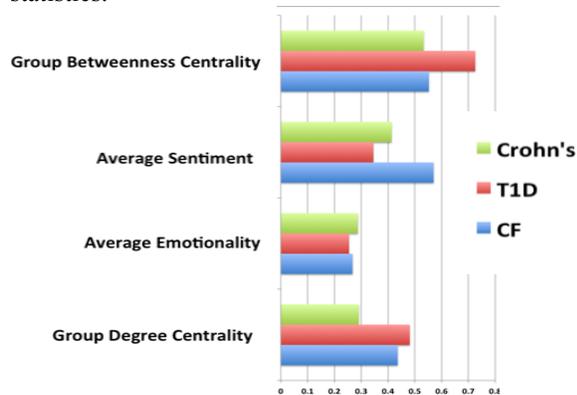

Figure 2: Key network metrics of 3 disease groups on Facebook

The T1D Facebook pages are the most centralized and have the highest density. This suggests that a few people might dominate the discussion, also acting as bridges between the different T1D groups – this is also visible in the graph in figure 1. The largest Crohn's and Cystic Fibrosis groups are operated by official patient organizations, which might lead to wider and less centralized communication with people sticking to communicating on the same page. T1D has the most negative sentiment on Facebook, while Crohn's is the most emotional, and Cystic Fibrosis posts are using the most complex language.

## ANALYZING TWITTER

Figure 3 illustrates the retweet network for the week Nov. 23 to Nov. 30, 2014. Each node in the network is a person, a connecting line means that one actor mentioned another one in a tweet, or retweeted a tweet by the other actor. As figure 3 shows, most tweets are made into the void, i.e. they do not trigger any reaction. Each of the three disease twitterers has a connected component, which is most dense for T1D, similar to the Facebook pages. There is also some people acting as connectors between the different disease groups, tweeting about two different diseases. The most central tweeters are a mix of individual patient activists, healthcare vendors, and patient organizations.

As figure 4 illustrates, Crohn's patients have the most negative sentiment, and the highest emotionality on Twitter. Crohn's tweeters are the most responsive (lowest ART), and also have the most similar tweeting contribution pattern (lowest variance in contribution index AWVCI) (Gloor et al. 2014).

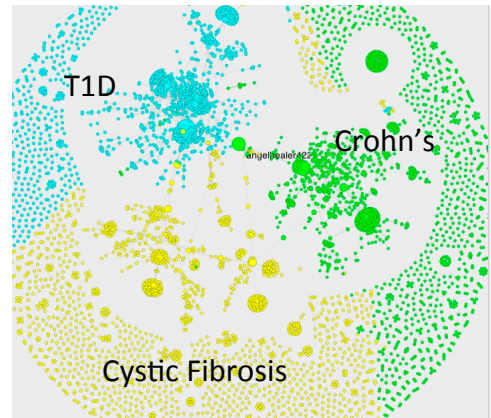

Figure 3: Twitter Network about 3 chronic diseases

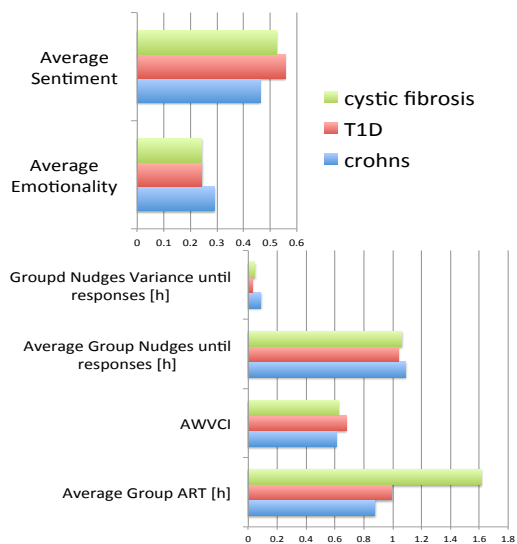

Figure 4: Key network metrics of 3 disease groups on Twitter

## CONCLUSIONS

In this early study we have shown that stakeholder's in Crohn's, which as a disease might appear "out of the blue" in the life of a patient are more emotional and negative than patients of Cystic Fibrosis, who have the disease since birth and are focused on creating and maintaining a long-term survival ecosystem.